\documentclass[aps,a4paper,oneside,twocolumn,superscriptaddress]{revtex4}
\usepackage[latin1]{inputenc}
\usepackage[]{natbib}
\usepackage{graphicx}
\usepackage{dcolumn}
\usepackage{amsmath}
\begin{document}
\title{Analytical Results for the Statistical Distribution Related to Memoryless Deterministic Tourist Walk: Dimensionality Effect and Mean Field Models}
\author{
\firstname{César} Augusto Sangaletti \surname{Terçariol}} 
\email{cesartercariol@gmail.com}
\affiliation{Centro Universitário Barão de Mauá \\
             Rua Ramos de Azevedo, 423 \\ 
             14090-180, Ribeirão Preto, SP, Brazil }
\affiliation{Faculdade de Filosofia, Ciências e Letras de Ribeirão Preto,
Universidade de São Paulo \\ Av. Bandeirantes, 3900 \\ 14040-901, 
Ribeirão Preto, SP, Brazil.}

\author{\firstname{Alexandre} Souto \surname{Martinez}}
\email{asmartinez@ffclrp.usp.br}
\homepage{http://fisicamedica.com.br/martinez/}
\affiliation{Faculdade de Filosofia, Ciências e Letras de Ribeirão Preto,
Universidade de São Paulo \\ Av. Bandeirantes, 3900 \\ 14040-901, 
Ribeirão Preto, SP, Brazil.}
\date{\today}

\begin{abstract}
Consider a medium characterized by $N$ points whose coordinates are randomly generated by a uniform distribution along the edges of a unitary $d$-dimensional hypercube.
A walker leaves from each point of this disordered medium and moves according to the deterministic rule to go to the nearest point which has not been visited in the preceding $\mu$ steps (deterministic tourist  walk). 
Each trajectory generated by this dynamics has an initial non-periodic part of $t$ steps (transient) and a final periodic part of $p$ steps (attractor).
The neighborhood rank probabilities are parameterized by the normalized incomplete beta function $I_d = I_{1/4}[1/2,(d+1)/2]$.
The joint distribution $S_{\mu,d}^{(N)}(t,p)$ is relevant, and the marginal distributions previously studied are particular cases.
We show that, for the memory-less deterministic tourist walk in the euclidean space, this distribution is: $S_{1,d}^{(\infty)}(t,p) = [\Gamma(1+I_d^{-1}) (t+I_d^{-1})/\Gamma(t+p+I_d^{-1})] \delta_{p,2}$, where $t=0,1,2,\ldots,\infty$, $\Gamma(z)$ is the gamma function and $\delta_{i,j}$ is the Kronecker's delta.
The mean field models are random link model, which corresponds to $d \rightarrow \infty$, and random map model which, even for $\mu = 0$, presents non-trivial cycle distribution [$S_{0,rm}^{(N)}(p) \propto p^{-1}$]: $S_{0,rm}^{(N)}(t,p) = \Gamma(N)/\{\Gamma[N+1-(t+p)]N^{t+p}\}$.  
The fundamental quantities are the number of explored points $n_e=t+p$ and $I_d$.
Although the obtained distributions are simple, they do not follow straightforwardly and they have been validated by numerical experiments.
\end{abstract}
\keywords{deterministic walk, tourist walk, random media, joint distribution, transient time distribution, attractor period distribution, random link model, random map model, extremum statistics, system dimensionality}

\maketitle


\section{Introduction}
\label{sec:introd}

The random walks in regular and disordered media are a very explored subject, capable to model several phenomena, in particular the transport problems~\cite{fisher:1984, barkema:2001}.
On the other hand, deterministic walks in disordered media are less explored and their behaviors are not completely understood yet.

Here we consider deterministic walks in disordered media.
The disordered medium is characterized by $N$ points whose coordinates are randomly generated, according to a uniform distribution, along the unitary edges of a $d$-dimensional hypercube.
A walker leaves from each point of this disordered medium and moves according to the deterministic rule to go to the nearest point which has not been visited in the last $\mu$ steps.
The quantity $\mu$ is called  memory and represents the required time (in number of steps) to the regeneration of the visited sites (refractory time).
The trajectory generated by this dynamics (deterministic tourist walk) has an initial non-periodic part of $t$ steps, called transient, and ends in a stable cycle of period $p$ steps (attractor) where  the same sites are visited in the same order. 
Although the dynamic may be simply stated, it has a complex behavior, with non-trivial results for $\mu \ge 2$~\cite{lima_prl2001, stanley_2001, kinouchi:1:2002, lima_phd, boyer_2004}. 
This rule may be relaxed, allowing the walker to visit the nearest sites with greater probabilities than the furthest ones (stochastic tourist walk)~\cite{risaugusman:1:2003, martinez:1:2004}.
Other than suggesting a possible mechanism for migration~\cite{lima_prl2001} this walk has been applied to thesaurus characterization~\cite{kinouchi:1:2002} and monkey movimentation~\cite{boyer_2004}.  

Depending on the system dimensionality $d$ and the memory $\mu$, several situations may be considered.
The simplest case is the lazy tourist ($\mu = 0$).
The walker remains caught in the initial site.
Thus all sites of the system are attractors of unitary period.
The joint distribution is $S_{0,d}(t,p) = \delta_{t,0} \delta_{p,1}$.
Although this situation is trivial, its extension to the stochastic tourist walk~\cite{martinez:1:2004} and for the random map model~\cite{derrida:2:1997} is interesting because of its analycity (glass transition in the former and non-trival cycle distribution in the latter). 
For $\mu = 1$, the tourist has to leave the site where he/she is and moves to the nearest site among the remainders.
Here the tourist knows, at each step, only the nearest site from his/her present position, but he/she  does not recall any of the sites visited. 
In this case, the trajectories always end in two sites, which are mutually nearest neighbors (couple).
Therefore, the period distribution is $S_{1,d}(p)  =  \delta_{p,2}$, but the transient distribution is not trivial and will be addressed here. 
The stochastic tourist walk with $\mu =1$ has been investigated in Ref.~\cite{risaugusman:1:2003}.
The cases $\mu \ge 2$ (treat with the quantity $\tau = \mu -1$) drastically differ from the preceding ones.
Even for $\mu = 2$, it is possible to obtain trajectories with long transients and periods.~\cite{lima_phd,stanley_2001,kinouchi:1:2002}.

The higher the euclidean space dimensionality is, the weaker are the correlations (as triangular inequality, for instance) among distances.
When $d \rightarrow \infty$, these correlations may be neglected and the distances between the sites may be considered as independent random variables.
Only the back/forwards symmetry $D_{i,j} = D_{j,i}$ is preserved.
This is the random link model (RLM), originally proposed by Mézard and Parisi~\cite{mezard:1986} and lately explored by Percus and Martin~\cite{percus:1999}.

If the back/forwards distances are different, the symmetry $D_{i,j} = D_{j,i}$ is not preserved and all the $N(N -1)$ distances are random independent variables.  
In this case, the only reminiscent characteristic related to previous model is the null distance from a site to itself ($D_{i,i}=0, \forall i$  and $\mu = 1$).
A variant of the asymmetric random link studied by Derrida and Flyvbjerg~\cite{derrida:2:1997}, called random map model (RMM), consists to eliminate the restriction $D_{i,i}=0$, allowing the distance from a site to itself not to vanish.
This distance may represent a cost for the tourist remains in a certain site, for instance.
Unlike previous models, even for $\mu=0$, the dynamics for this model gives rise to a complex cycle period distribution. 
This walk corresponds to the mean field approximation for the Kauffman's networks~\cite{kauffman:1969}.

The main objective of this paper is to generalize the geometric distribution and analytically obtain the probability joint distribution for transient time and attractor period for a $\mu = 1$ deterministic tourist walk. 
Also, we study how these distributions are affected by the dimensionality and the border/finite-size  effects. 
We stress that $I_d$ and the number of visited sites $n_e$ are the relevant quantities of the problem. 
To solve this problem we have adopted the following strategy. 
Firstly, the solutions have been obtained in the limiting dimensionalities, i.e., $d \rightarrow \infty$ and $d=1$, and then the distribution for finite $d$ has been inferred through a generalization of the geometrical distribution.
All obtained results have been numerically validated.

This article is divided as it follows:
In Sec.~\ref{sec:prob_viz}, we present the parameterization of Cox's equation by $I_d$. 
In Sec.~\ref{sec:rand_link}, we analytically determine the transient distribution for the RLM for arbitrary $N$.
An analogy to the geometric distribution is then established.
The subsistence and capture probabilities through the walk are defined, which will be the standard interpretation to treat the considered models.
We then focus on the one-dimensional (1D) systems.
With a simple modification in the algebraic formulation, the transient distribution for the infinite medium is obtained.
Finally, we generalize the obtained results to systems with arbitrary dimensionalities and numerically show their validity. 
In Sec.~\ref{sec:rand_map}, the joint distribution of the transient time and attractors period for the RMM is obtained.
Final considerations are presented in Sec.~\ref{sec:concl}, where the role of $I_d$ and $n_e = t+p$ to the joint distributions is stressed. 
In Appendix~\ref{apendice1}, some special functions are recalled amd in Appendix~\ref{apendice2} the cumulative distribution is obtained in terms of the subsistence probabilities. 

\section{Reflexive Neighbors}
\label{sec:prob_viz}

In a Poissonic process of dimensionality $d$, the probability that an arbitrary event is the $m^{\mbox{th}}$ nearest neighbor of its own $n^{\mbox{th}}$ nearest neighbor is given by Cox's equation~\cite{cox}:
\begin{eqnarray}
\nonumber
P_{m,n}^{(d)} & = & \frac{\left(I_d^{-1}+1\right)^{-(m+n)}}{1-I_d} \\ 
              &   & \sum_{j=1}^{\scriptsize{\mbox{min}}(m,n)} 
                    \frac{\left(I_d^{-2}-1\right)^j \Gamma(m+n-j)}
                         {\Gamma\left(j\right)\Gamma(m+1-j)\Gamma(n+1-j)} \; .
\label{eq:cox}
\end{eqnarray}
The medium dimensionality is implicitly considered in: 
\begin{eqnarray}
I_d & = & I_{1/4} \left(\frac{1}{2}, \frac{d+1}{2}\right) \approx 1 - \frac{e^{d/8}}{\sqrt{\pi d / 8}} \; ,
\label{eq:Id}
\end{eqnarray}
(denoted by $p$ in Ref.~\cite{cox}) which is written in terms of the normalized incomplete beta function (Eq.~\ref{eq:beta_incompl} and~\ref{eq:gama_aprox}). 
As shown in Appendix~\ref{apendice1}, the approximation is justified for $d \gg 1$. 
This is the relevant quantity that parameterizes the transient time distributions studied here. 
This quantity suggests a characteristic dimensionality $(d_0=8)$ from which the effects related to the dimensionality of the system can be neglected.

In particular, for $m=n$, Eq.~\ref{eq:cox} gives the $n$-order reflexive neighbors probability (Dacey's equation~\cite{dacey}):
\begin{eqnarray}
P_{n}^{(d)}
& = & \frac{\left(I_d^{-1}+1\right)^{-2n}}{1-I_d} \sum_{j=1}^n \frac{\left(I_d^{-2}-1\right)^j\Gamma(2n-j)}{\Gamma(j) \Gamma^2(n+1-j)} \; .
\label{eq:dacey}
\end{eqnarray}
The 1-order reflexive neighbors probability represents the probability of finding the medium attractors for $\mu=1$ walkers, and Eq.~\ref{eq:dacey} reduces to:
\begin{eqnarray}
\label{eq:p1d}
P_1^{(d)} & = & \frac{1}{1+I_d} \; ,
\end{eqnarray}
and $P_1^{(d)}  =  1/(2- e^{-d/8}/\sqrt{\pi d/8})$, when $d \gg 1$. 
The quantity $P_1^{(d)}$ can also be interpreted as the null transient time probability$S_{1,d}(t=0) = P_1^{(d)}$ in the $\mu =1$ tourist walk. 
The value of $P_1^{(d)}$ ranges from $3/4$, for $d=0$, and diminishes monotoly, converging exponentially to the asymptotic value  $1/2$, as $d \rightarrow \infty$.~\cite{tercariol_msc} 
Notice that Eq.~\ref{eq:p1d} produces results for any real value $d$, which opens the possibility to the interpretation to non-integer dimensionality (fractals). 
Curiously, the analytical continuation for $d=0$ leads to $P_1^{(0)} = 3/4$ and not $P_1^{(0)} = 1$ as one would possibly expect.

\section{Transient Time Distribution} 
\label{sec:rand_link}

Here we obtain the treansient time distribution for $\mu = 1$ tourist walk. 
We start obtaining analytical results for the RLM (since the site distances are independent random variables), then for 1D systems, and finally the result is inferred to an arbitrary dimensionality systems. 

\subsection{Random Link Model}

The RLM represents a mean field model for the euclidean space high dimensionality limit.
Consider $N$ independent continuous random variables $X_1$, $X_2$, $\ldots$, $X_N$, with probability density functions (pdf's) $f_1(x_1)$, $f_2(x_2)$, $\ldots$, $f_N(x_N)$, respectively.
The random variable $Y= \min\{X_1, X_2, \ldots, X_N\}$ pdf $g(y)$ can be determined as follow~\cite{chapman:2002}:
The condition for $Y$ to assume a given value $y$ is that at least one of the variables $X_1$, $X_2,$ $\ldots$, $X_N$ is equal to $y$ and all of the remainders are greater or equal to $y$:
$g(y) = \sum_{i=1}^N f_i(y) \prod_{j \, (\ne i) = 1}^N \int_y^\infty \mbox{d}x_j f_j(x_j)$.

If all variables $X_1$, $X_2$, $\ldots$, $X_N$ have the same pdf $f(x)$, then: $g(y)  =  N f(y) [1 - F(y) ]^{N-1}$,  where
$F(y) = \int_{-\infty}^y \mbox{d}x f(x)$ is the $f(x)$ cumulative function. 
In particular, if $f(x)$ is a uniform between 0 and 1 then $g(y) = N\left(1-y\right)^{N-1}$, with $y$ varying from $0$ to $1$.

In the high dimensionality limit, the distances are independent random variables, but what is the  pdf?
In fact, it does not matter to the tourist walk. 
Let us see the reason. 
For a given random variable $X$, following the pdf $f(x)$, one can define another random variable $Y=h(X)$, with some aimed pdf $g(Y)$ simply imposing the condition that their cumulative distributions are equal, $G(Y) = F(X)$.
If $g(y)>0$, for all possible values of $Y$, then the cumulative function $G$ is a bijection, and the definition $Y = G^{-1}[F(X)]$ leads to $h = G^{-1} \circ F$.
Thus, if the distances $D_{i,j}$ are particular values of a random variable $X$, with pdf  $f(x)$, it is then possible to obtain the random variable $Y=h(X)$ with uniform pdf $g(y)=1$ in the interval $[0; 1]$.
In this case, the cumulative function $G$ is the identity function, so its inverse is the identity function too.
Hence, $Y =  F(X) = \int_0^X \mbox{d}x \, f(x)$.
As $f(x) > 0, \forall x$, one has that $Y$ is a strictly crescent function of $X$.
Therefore, if the distances $D_1$ and $D_2$ obey the relation $D_1 < D_2$ in the $X$ metrics, their correspondent values in the metrics $Y$ will also do.
For the deterministic tourist walk, both metrics yield the same trajectories, since in each step, it is not important the distance to be run, but only the neighborhood rank.

The transient time probability distribution for a deterministic walk, with $\mu = 1$, can be obtained for the RLM by noting that: \emph{i)} the distance matrix is symmetric, \emph{ii)} 
the distance between the sites are independent random variables and uniformly distributed in the interval from $0$ to $1$, and \emph{iii)} the walk distance decreases at each step, until the tourist enters a 2-cycle.

The walk distance $x_1$ in the first step is the minimum of the $N-1$ independent random variables.
In the second step, only $N-2$ new independent distances are available (because of the metrics symmetry, the distance to the first site has already been evaluated).
Thus, for a trajectory with a transient size $t \le N-2$, it is necessary and sufficient that $x_{t+1} < x_t < \ldots < x_2 < x_1$ and that $x_{t+2}>x_{t+1}$, where $x_j$ represents the minimum of the distances to the $N-j$ sites not visited up to the $j^{\mbox{th}}$ step.
This leads to the following transient time distribution:
$S_{1,rl}^{(N)}(t)
 =  \int_0^1\mbox{d}x_1(N-1)(1-x_1)^{N-2} \prod_{j=2}^{t+1} \int_0^{x_{j-1}}\mbox{d}x_j(N-j)(1-x_j)^{N-j-1}  \int_{x_{t+1}}^1\mbox{d}x_{t+2}(N-t-2)(1-x_{t+2})^{N-t-3}$. 
Consider $a_i = N - (t + 2) + i$, with $i = 0, 1, \ldots, t+1$, so that: 
\begin{eqnarray}
\nonumber
S_{1,rl}^{(N)}(t)
& = & \left[ \prod_{i=0}^{t+1} a_i \right] \int_0^1\mbox{d}x_1(1-x_1)^{a_{t+1}-1} \\ \nonumber
&   & \prod_{j=2}^{t+1} \int_0^{x_{j-1}} \mbox{d}x_j(1 - x_j)^{a_{t+2-j}-1} \\
&   & \int_{x_{t+1}}^1\mbox{d}x_{t+2}(1-x_{t+2})^{a_0-1} \; .
\label{eq:distr_RL_prod_int}
\end{eqnarray}
The Fig.~\ref{eq:distr_RL_prod_int} schema shows the calculation of the  $t+2$ chained integrals of the Eq.~\ref{eq:distr_RL_prod_int} for some particular values of $t$.
Summing these terms (and multiplying the sum by the produtory of the Eq.~\ref{eq:distr_RL_prod_int}), one obtains generically:
$ S_{1,rl}^{(N)}(t) = a_1\left[\prod_{j=2}^t a_j / \sum_{k=j}^{t+1} a_k \right]/\sum_{i=0}^{t+1} a_i$,
where $\sum_{i=0}^{t+1} a_i = [N-(t+3)/2] (t+2)$ and $\sum_{k=j}^{t+1} a_k  = [N- (t+3-j)/2](t+2-j)$. 
Calling $k = t+2-j$ leads to:
\begin{eqnarray}
\label{eq:distr_RL_tam_finito}
\frac{S_{1,rl}^{(N)}(t)}{S_{1,rl}^{(\infty)}(t)} & = & \frac{N-t-1}{N-(t+3)/2} \prod_{k=2}^t \frac{N-k}{N-(k+1)/2} \; ,
\end{eqnarray}
where $t=0,1,2,\ldots,N-2$ and the transient time distribution of the RLM in the thermodynamic limit ($N \gg 1$) is
\begin{eqnarray}
\label{eq:distr_RL_lim_termo}
S_{1,rl}^{(\infty)}(t) & = & \frac{t+1}{(t+2)!} \; ,
\end{eqnarray}
which leads to $\overline{t} = e-2$ and $\sigma_t^2 = e(1-\overline{t})$. 

\begin{figure}
\begin{center}
\includegraphics[angle=90,scale = 1]{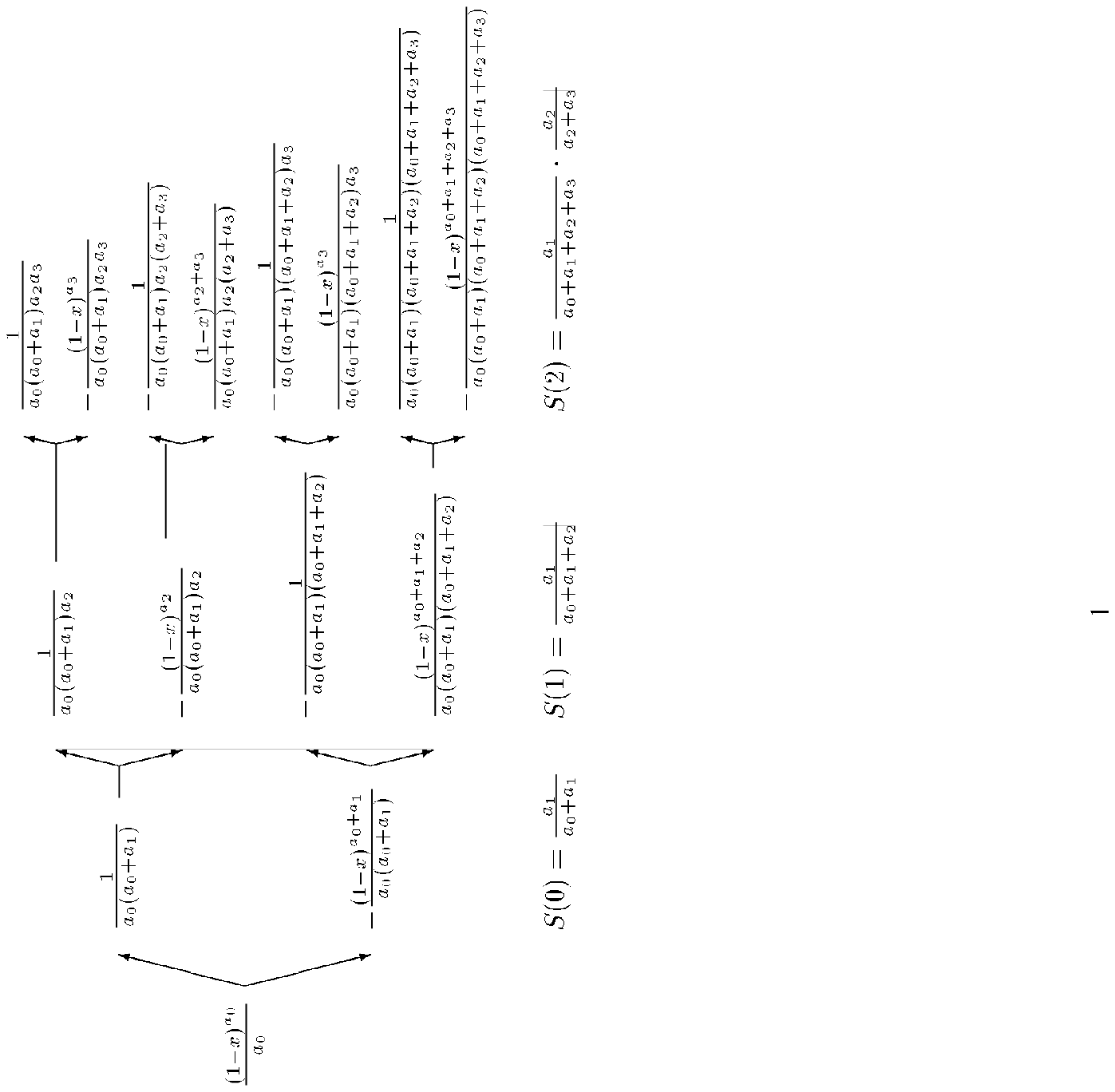}
\vspace{1cm}
\caption{Chained integral calculation of Eq.~\ref{eq:distr_RL_prod_int}. 
In each integration step, the number of integrals is doubled.  
The summations are indicated in the bottom line.}
\label{fig:calc_int_encad_RL}
\end{center}
\end{figure}

\begin{figure}[htb]
\begin{center}
\includegraphics[angle=-90,width = \columnwidth]{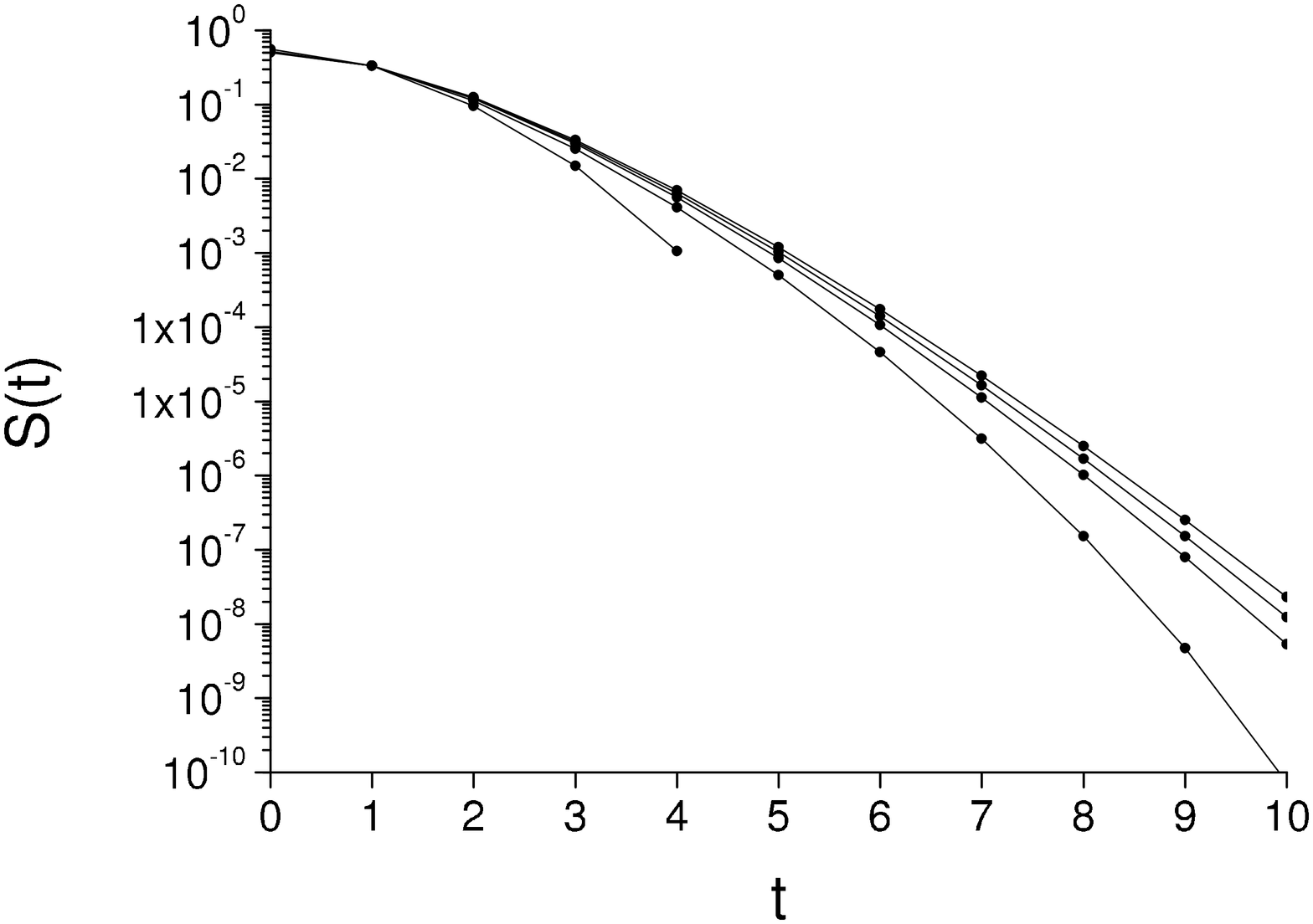}
\caption{Transient time distribution for the random link model. 
From bottom to top, the curves refer respectively to $N=6, 12, 25, 50, 100, 200$ and to limit $N \gg 1$.}
\label{fig:RL_efeito_tam_finito}
\end{center}
\end{figure}

Fig.~\ref{fig:RL_efeito_tam_finito} shows the transient time distribution for some values of $N$ and the convergence to the thermodynamic limit.
Observe that, in particular, $S_{1,rl}^{(\infty)}(0) = 1/2 = P_1^{(\infty)}$ (Eq.~\ref{eq:p1d}).
An immediate property is that: $S_{1,rl}^{(N)}(1) = 1/3$.

To measure the finite size effect for a given $N$, the mean square error:  $E_{1,rl}(N) = \sum_{t=0}^{\infty} [ S_{1,rl}^{(N)}(t) - S_{1,rl}^{(\infty)}(t)]^2/N $ has been considered.
Numerically, one verifies that, for $N > 10$, $E_{1,rl}$ is a power law~\cite{tercariol_msc}: $E_{1,rl}(N) \approx 0,08611 N^{-3}$.


The probability the tourist enters an attractor is not constant along the walk, but it can be analytically expressed as a function of the trajectory step.
Let us start with the thermodynamic limit.
The cumulative distribution is $F_{1,rl}^{(\infty)}(t) = 1 - 1/(t+2)!$ and 
Eq.~\ref{eq:distr_RL_lim_termo} can be alternatively written as:
$ S_{1,rl}^{(\infty)}(t)  =  [ 1- F_{1,rl}^{(\infty)}(t-1)] (t+1)/(t+2) = (t+1)/[(t+1)!(t+2)]$, where $1/(t+1)!$ represents the probability of the walker subsisting (not entering an attractor) in the $t$ initial steps, and $(t+1)/(t+2)$ represents the probability of the walker being captured (entering an attractor) in the $(t+1)^{\mbox{th}}$ step. 
This leads to:
$S_{1,rl}^{(\infty)}(0) =  1/2$, 
$S_{1,rl}^{(\infty)}(1) =  1/2 \times 2/3$, 
$S_{1,rl}^{(\infty)}(2) =  1/2 \times 1/3 \times 3/4$, 
$S_{1,rl}^{(\infty)}(3) =  1/2 \times 1/3 \times 1/4 \times 4/5 \; \cdots$. 
Making an analogy to the geometric distribution, Eq.~\ref{eq:distr_RL_lim_termo} can also be written as: 
\begin{equation}
S_{1,rl}^{(\infty)}(t) = [1-q_{1,rl}^{(\infty)}(j)(t+1)] \prod_{j=1}^t q_{1,rl}^{(\infty)}(j) \; ,
\label{eq:rl_geom} 
\end{equation}
where the success and, consequently, the failure probabilities depend on the extraction stage.  
The subsistence probability up to the $j^{\mbox{th}}$ step is: 
\begin{equation}
q_{1,rl}^{(\infty)}(j) = \frac{1}{j+1} \; .
\label{eq:q_rl}
\end{equation} 
A similar reasoning can be applied to finite size systems.
With some manipulations, Eq.~\ref{eq:distr_RL_tam_finito} can be expressed in the form
$S_{1,rl}^{(N)}(t)  =  [1- (N-t-2)/\{[N-(t+1)/2-1] (t+2)\} ] \prod_{k=1}^t (N-k-1)/[( N- k/2 -1) (k+1)]$, which allows us to obtain the finite size correction factor for the subsistence probability:
\begin{equation}
\frac{q_{1,rl}^{(N)}(j)}{q_{1,rl}^{(\infty)}(j)}  =  \frac{N-j-1}{N-j/2-1} \; .
\end{equation}
In terms of the subsistence probabilities,  Eq.~\ref{eq:distr_RL_tam_finito} can be rewritten as:
$S_{1,rl}^{(N)}(t)  =  [1-q_{1,rl}^{(N)}(t+1)] \prod_{j=1}^t q_{1,rl}^{(N)}(j)$ and the cumulative distribution is $F_{1,rl}^{(N)}(t) = 1 - \prod_{j=1}^{t+1} q_{1,rl}^{(N)}(j)$ can be determined according to the Appendix~\ref{apendice2}.
Observe that using $F_{1,rl}^{(N)}(t)$, one can easily verify that the transient distribution for finite size is also normalized $\sum_{k=0}^{N-2} S_{1,rl}^{(N)}(k)  = 1$, since $q_{1,rl}^{(N)}(N-1) = 0$.

\subsection{One-Dimensional Systems}
\label{sec:sist_unidim}

As the RLM represents the upper limit for the dimensionality ($d \rightarrow \infty$) in the euclidean space, the 1D model represents the lower limit, which is the easiest to analyze for finite dimensionality systems. 
In this Section, the transient time distribution for 1D systems in the thermodynamic limit is analytically obtained.
The demonstration considers a semi-infinite medium, which establishes a surprising connection with the RLM.
With a simple modification in the previous model, the transient time distribution for the infinite medium can be obtained. 

\subsubsection{Semi-infinite medium}

The analysis of the semi-infinite disordered medium aims to: $i$) develop the calculation kernel for the transient distribution in the infinite medium, $ii$) reveal a non-trivial equivalence between 1D systems and the RLM, and $iii$) evaluate the edge effect.

The semi-infinite medium can be thought as a set of uncountable points randomly and uniformly distributed on a line-segment, with a mean density of $r$ points per unitary length.
Consider site $s_1$ placed at the origin of the line-segment.

This model is described by a 1D Poissonic process.
The sites can be viewed as events, which occur randomly as time flows.
The distances between consecutive sites follow an exponential pdf:  
$f(x) = re^{-rx}$ for $x \ge 0$ and $f(x) = 0$ otherwise.

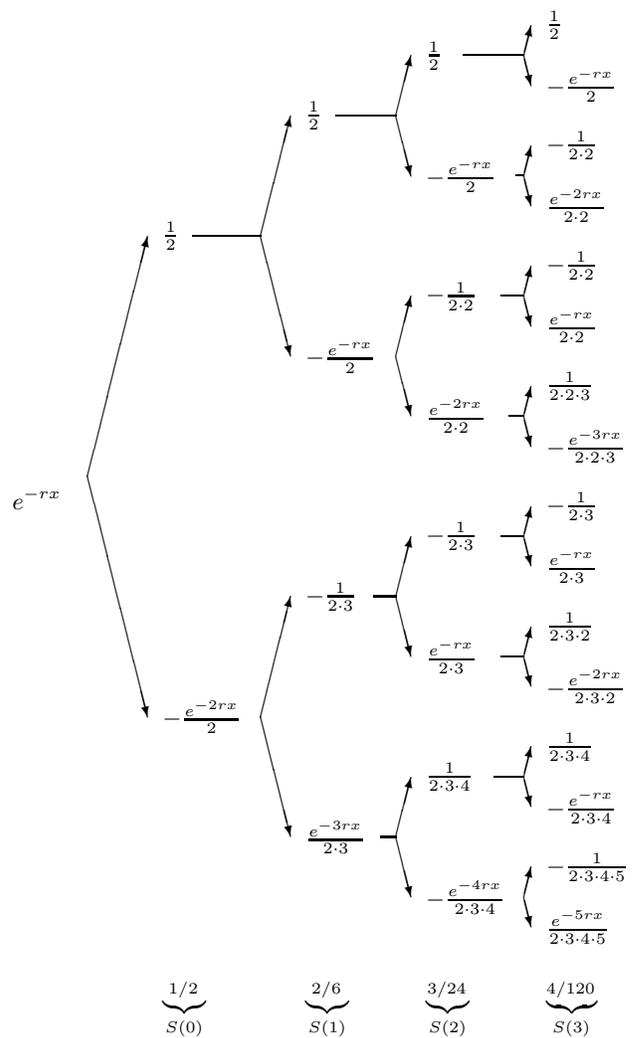
\begin{figure}[htb]
\begin{center}
\setlength{\unitlength}{1mm}
\begin{picture}(88,138)
\put(0,72){\makebox(0,0)[l]{$e^{-rx}$}}

\put(10, 75){\vector(1, 4){8}}
\put(10, 75){\vector(1,-4){8}}
\put(20,107){\makebox(0,0)[l]{$\frac{1}{2}$}}
\put(20, 43){\makebox(0,0)[l]{$-\frac{e^{-2rx}}{2}$}}

\put(20, 4){\makebox(0,0)[l]{$\underbrace{_{1/2}}_{S(0)}$}}

\put(33,107){\line(-1, 0){9}}
\multiput(33,43)(0,64){2}{\vector(1, 4){4}}
\multiput(33,43)(0,64){2}{\vector(1,-4){4}}
\put(39,123){\makebox(0,0)[l]{$\frac{1}{2}$}}
\put(39, 91){\makebox(0,0)[l]{$-\frac{e^{-rx}}{2}$}}
\put(39, 59){\makebox(0,0)[l]{$-\frac{1}{2\cdot3}$}}
\put(39, 27){\makebox(0,0)[l]{$\frac{e^{-3rx}}{2\cdot3}$}}

\put(39, 4){\makebox(0,0)[l]{$\underbrace{_{2/6}}_{S(1)}$}}

\put(51,123){\line(-1, 0){8}}
\put(51, 59){\line(-1, 0){3}}
\put(51, 27){\line(-1, 0){2}}
\multiput(51,27)(0,32){4}{\vector(1, 4){2}}
\multiput(51,27)(0,32){4}{\vector(1,-4){2}}
\put(55,131){\makebox(0,0)[l]{$\frac{1}{2}$}}
\put(55,115){\makebox(0,0)[l]{$-\frac{e^{-rx}}{2}$}}
\put(55, 99){\makebox(0,0)[l]{$-\frac{1}{2\cdot2}$}}
\put(55, 83){\makebox(0,0)[l]{$\frac{e^{-2rx}}{2\cdot2}$}}
\put(55, 67){\makebox(0,0)[l]{$-\frac{1}{2\cdot3}$}}
\put(55, 51){\makebox(0,0)[l]{$\frac{e^{-rx}}{2\cdot3}$}}
\put(55, 35){\makebox(0,0)[l]{$\frac{1}{2\cdot3\cdot4}$}}
\put(55, 19){\makebox(0,0)[l]{$-\frac{e^{-4rx}}{2\cdot3\cdot4}$}}

\put(55, 4){\makebox(0,0)[l]{$\underbrace{_{3/24}}_{S(2)}$}}

\put(68,131){\line(-1, 0){8}}
\put(68,115){\line(-1, 0){1}}
\put(68, 99){\line(-1, 0){3}}
\put(68, 83){\line(-1, 0){2}}
\put(68, 67){\line(-1, 0){3}}
\put(68, 51){\line(-1, 0){3}}
\put(68, 35){\line(-1, 0){4}}
\multiput(68,19)(0,16){8}{\vector(1,4){1}}
\multiput(68,19)(0,16){8}{\vector(1,-4){1}}
\put(71,135){\makebox(0,0)[l]{$\frac{1}{2}$}}
\put(71,127){\makebox(0,0)[l]{$-\frac{e^{-rx}}{2}$}}
\put(71,119){\makebox(0,0)[l]{$-\frac{1}{2\cdot2}$}}
\put(71,111){\makebox(0,0)[l]{$\frac{e^{-2rx}}{2\cdot2}$}}
\put(71,103){\makebox(0,0)[l]{$-\frac{1}{2\cdot2}$}}
\put(71, 95){\makebox(0,0)[l]{$\frac{e^{-rx}}{2\cdot2}$}}
\put(71, 87){\makebox(0,0)[l]{$\frac{1}{2\cdot2\cdot3}$}}
\put(71, 79){\makebox(0,0)[l]{$-\frac{e^{-3rx}}{2\cdot2\cdot3}$}}
\put(71, 71){\makebox(0,0)[l]{$-\frac{1}{2\cdot3}$}}
\put(71, 63){\makebox(0,0)[l]{$\frac{e^{-rx}}{2\cdot3}$}}
\put(71, 55){\makebox(0,0)[l]{$\frac{1}{2\cdot3\cdot2}$}}
\put(71, 47){\makebox(0,0)[l]{$-\frac{e^{-2rx}}{2\cdot3\cdot2}$}}
\put(71, 39){\makebox(0,0)[l]{$\frac{1}{2\cdot3\cdot4}$}}
\put(71, 31){\makebox(0,0)[l]{$-\frac{e^{-rx}}{2\cdot3\cdot4}$}}
\put(71, 23){\makebox(0,0)[l]{$-\frac{1}{2\cdot3\cdot4\cdot5}$}}
\put(71, 15){\makebox(0,0)[l]{$\frac{e^{-5rx}}{2\cdot3\cdot4\cdot5}$}}

\put(71, 4){\makebox(0,0)[l]{$\underbrace{_{4/120}}_{S(3)}$}}

\end{picture}
\caption{Calculation of chained integrals of Eq.~\ref{eq:distr_E1_cac}. Each integration step doubles the number of integrals.}
\label{fig:calc_int_encad_E1}
\end{center}
\end{figure}

The transient time distribution for a deterministic walk of a tourist who leaves from the origin of the system, with $\mu=1$, can be determined by noting that the run distance decreases each step.
For a trajectory with a $t$ transient steps, it is necessary and sufficient that $x_{t+1} < x_t < \ldots < x_2 < x_1$ and $x_{t+2}>x_{t+1}$, where $x_j$ represents the run distance in the $j^{th}$ step.
This leads to: 
\begin{eqnarray}
\nonumber
S(t) & = & \int_0^{\infty}\mbox{d}x_1re^{-rx_1} \prod_{j=2}^{t+1} \int_0^{x_{j-1}}\mbox{d}x_jre^{-r x_j} \\ 
     &   & \int_{x_{t+1}}^{\infty}\mbox{d}x_{t+2}re^{-rx_{t+2}} = \frac{t+1}{(t+2)!} \; , 
\label{eq:distr_E1_cac}
\end{eqnarray} 
where the schema of the Fig.~\ref{fig:calc_int_encad_E1} ilustrates the calculation of theses integrals.

This expression is equal to Eq.~\ref{eq:distr_RL_lim_termo}, the thermodynamic limit for the RLM.
It is also interesting to observe that the medium mean density ($r$ points by an unitary length) does not interfere in the transient time distribution, it only affects the distance the tourist runs each step.

It may seem strange, at first look; that two models that represent opposite limits,  concerning the dimensionality, present the same expression for the transient time distribution.
This equivalence can be understood noting that although there exist strong correlations in the distance matrix for 1D systems, the distances that really matter for $\mu=1$ tourist are those ones between consecutives sites, and in fact, these distances are independent variables.
Numerical simulations have revealed that this equivalence does not hold for $\mu=2$, because in this case second neighbor distances are also important~\cite{tercariol_msc}.

\subsubsection{Infinite Medium}

With a simple modification in the previous model, one is able to obtain analytically the transient time distribution for a ``left-and-right hand unlimited'' medium.
Consider the tourist leaves from site $s_1$ of the infinite medium.
For a transient time $t$, as mentioned before, the only additional condition is that the distance $x_0$ (between the sites $s_0$ and $s_1$) must be greater than the distance $x_1$ (between the sites $s_1$ and $s_2$).
One must also multiply the expression by $2$, since the walk can start either to the left or righthand side. 
Applying these cosiderations to the Eq.~\ref{eq:distr_E1_cac}, one has:
$S_{1,1}^{(\infty)}(t)  =  2 \int_0^{\infty}\mbox{d}x_1re^{-rx_1} \int_{x_1}^{\infty}\mbox{d}x_0re^{-rx_0} \prod_{j=2}^{t+1} \int_0^{x_{j-1}}\mbox{d}x_jre^{-rx_j}  \times \int_{x_{t+1}}^{\infty}\mbox{d}x_{t+2}re^{-rx_{t+2}}$. 
The integral in $x_0$ is not linked with the other ones, and can be simply calculated. 
In this way, the schema of the Fig.~\ref{fig:calc_int_encad_E1} became a little different.
In the last integration stage, the term $-rx$ is added to the $e^{-rx}$, $e^{-2rx}$, $e^{-3rx}\; \ldots$ exponents.
Consequently, the denominators factors $2$, $3$, $4$, \ldots, from the last integral will be increased by a unity.
Hence:
$S_{1,1}^{(\infty)}(0)  =  2 \times 2/6$, 
$S_{1,1}^{(\infty)}(1)  =  2 \times 3/24$, 
$S_{1,1}^{(\infty)}(2)  =  2 \times 4/120$, 
$S_{1,1}^{(\infty)}(3)  =  2 \times 5/720$, 
$S_{1,1}^{(\infty)}(4)  =  2 \times 6/5040$, 
$S_{1,1}^{(\infty)}(5)  =  2 \times 7/40320$, $\ldots$ .
Generically, for $t=0,1,2,\ldots,\infty$, one has:
\begin{eqnarray}
\label{eq:distr_E1_meio_infinito}
S_{1,1}^{(\infty)}(t) & = & \frac{2(t+2)}{(t+3)!} \; .
\end{eqnarray}
In this way, we have analytically obtained the normalized transient time distribution for 1D $\mu=1$ systems with $N \gg 1$. 
This leads to: $\overline{t} = 2e-5$ and $\sigma_t^2 =  2e(2-\overline{t})-8$ .
Notice that $S_{1,1}^{(\infty)}(0) = 2/3 = P_1^{(1)}$ (Eq.~\ref{eq:p1d}).

Similarly to the RLM, the probability the tourist is trapped in an attractor at each walk step in the 1D case can also be analytically obtained.
The transient time cumulative distribution is:
$F_{1,1}^{(\infty)}(t) = \sum_{k=0}^t S_{1,1}^{(\infty)}(k) = 1-2/(t+3)!$ and the recursive form of  Eq.~\ref{eq:distr_E1_meio_infinito} is $S_{1,1}^{(\infty)}(t)  =  [ 1- F_{1,1}^{(\infty)}(t-1)] (t+2)/(t+3) = 2(t+2)/[(t+2)! (t+3)]$. 
So that: $S_{1,1}^{(\infty)}(0) = 2/3, S_{1,1}^{(\infty)}(1) = 1/3  \times 3/4, S_{1,1}^{(\infty)}(2) = 1/3 \times 1/4 \times 4/5, S_{1,1}^{(\infty)}(3) = 1/3 \times 1/4 \times 1/5 \times 5/6$.
In this way:
\begin{equation}
S_{1,1}^{(\infty)}(t)  =  [1-q_{1,1}^{(\infty)}(t+1)] \prod_{j=1}^t q_{1,1}^{(\infty)}(j) \; ,
\label{eq:1d_trans_geom}
\end{equation}
where the success and failure probabilities depend on the stage of the extraction:
\begin{equation}
q_{1,1}^{(\infty)}(j)  =  \frac{1}{j+2}
\label{eq:1_1d}
\end{equation}
represents the subsistence probability (not getting an attractor) up to the $j^{th}$ step.

The notable regularity in RLM and 1D models makes us to consider the subsistence probability as the proper quantity to the generalization for arbitrary dimensionality systems.
Contrasting to the RLM, we have numerically verified, for 1D systems, that the transient time distribution has a weak dependence on $N$.~\cite{tercariol_msc}

\subsection{Arbitrary Dimensionality}
\label{sec:dim_arbitr}

\begin{figure}[htb]
\begin{center}
\includegraphics[angle=-90,width = \columnwidth]{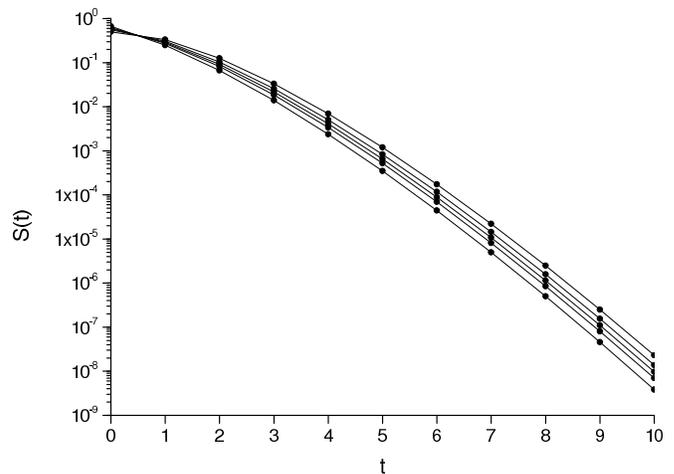}
\caption{Effect of the dimensionality on the transient distribution. 
From bottom up, the curves refer to $d=1,2,3,5$ and $\infty.$}
\label{fig:DimensionalidadeFinita}
\end{center}
\end{figure}

Here we present arguments which allow us to predict the analytical form of the transient time distribution for the $\mu = 1$ tourist walk in systems with an arbitrary dimensionality for $N \gg 1$.  
As shown in Fig.~\ref{fig:DimensionalidadeFinita}, numerical simulation results have revealed that the transient time distributions for arbitrary dimensionalities lies between the analytically obtained limiting distributions ($d=1$ and $d \rightarrow \infty$).

Comparing Eqs.~\ref{eq:1d_trans_geom} and~\ref{eq:rl_geom} one notices the same mathematical structure. 
Further, the null transient trajectory probability in a $d$-dimensional medium is (Eq.~\ref{eq:dacey}):
$S_{1,d}^{(\infty)}(0) = P_1^{(d)} = 1/(1+I_d)$. 
This is also the probability the tourist is captured in the first step. 
Thus, the subsistence probability in the first step is
$q_{1,d}^{(\infty)}(1) = 1 - P_1^{(d)} =  1/(1+I_d^{-1})$.
Comparing Eqs.~\ref{eq:q_rl} and~\ref{eq:1_1d}, we infer that the subsistence probability for each step of the trajectory for arbitrary dimensionalities is given by:
\begin{equation}
q_{1,d}^{(\infty)}(j)  =  \frac{1}{j+I_d^{-1}} \; .
\end{equation}

From these subsistence probabilities, it is possible to build a closed analytical expression for the transient time distribution for arbitrary dimensionality through the analogy with the geometric distribution: 
\begin{eqnarray}
S_{1,d}^{(\infty)}(t)
& = & \frac{\Gamma(1+I_d^{-1}) (t+I_d^{-1})}{\Gamma(t+2+I_d^{-1})} \; ,
\label{eq:distr_E_arbitr}
\end{eqnarray}
with $t=0,1,2,\ldots,\infty$, which leads to   $\overline{t} = e \left[ \Gamma(1+I_d^{-1}) - \Gamma(1+I_d^{-1},1) \right]$.
Although the obtained results have been based on a conjecture, numerical simulations have confirmed their  validity.

Consider the following remarks.
One can write this distribution in terms of $I_d$ and in terms of the number $n_e=t+p$ of explored sites, once $p=2$  remaining a $t$. 
The cumulative distribution can be obtained according to the Appendix~\ref{apendice2}:
$F_{1,d}^{(\infty)}(t) =  1 - \Gamma(1+I_d^{-1})/\Gamma(t+2+I_d^{-1})$.

\section{Joint Distribution in the Random Map Model}
\label{sec:rand_map}

The Derrida-Flyvbjerg random map~\cite{derrida:2:1997} is a mean field approximation for the Kauffman~\cite{kauffman:1969} network.
The map is built associating to each one of the $N$ sites a random site; and the movement rule is to go, at each time step, to the successor site.
Eventually a site may be its own successor. 
Even for $\mu=0$ and $\mu=1$ the model presents a non-trivial period distribution.
This model may be applied to situations where the concept of distance represents a cost, and the links are represented by a directed graph.
In the RMM, contrasting to the preceding studied cases, even with $\mu=0$, it is possible to obtain periods varying from $1$ to $N$.
Therefore, a joint distribution for the transient time and attractors period is required to completely describe the walk.

Consider that the tourist starts the walk at site $s_1$.
For the walk to have a transient $t=0$ and a period $p=1$ (i.e., consists on a single point), the site following $s_1$ must be $s_1$ itself.
Hence, the probability is $S_{0,rm}^{(N)}(0,1)  =  1/N$.
For the walk to have $t=1$ and $p=1$, the tourist must go to any site $s_2$  among the $N-1$ reminders, and in the following step to remain at $s_2$, leading to:  $S_{0,rm}^{(N)}(1,1)  =  (N-1)/N \times 1/N$. 
Thus, the transient time marginal distribution for $p=1$ is
$S_{0,rm}^{(N)}(t,1)  =  (N-1)/N \times (N-2)/N \cdots (N-t)/N \times 1/N$.
To obtain a $t=0$ transient and a $p = 2$ period, the walker must go to any of $N-1$ remaining sites and, in the second step, return to  $s_1$: $S_{0,rm}^{(N)}(0,2)  =  (N-1)/N \times 1/N$. 
For $t=1$ and $p=2$, the walker must go to any $N-1$ remaining site, in the second step go to any  of the $N-2$ reminders, and finally return to $s_2$: $S_{0,rm}^{(N)}(1,2)  =  (N-1)/N \times (N-2)/N \times 1/N$. 
The transient time marginal distribution for $p=2$ is: 
$ S_{0,rm}^{(N)}(t,1)  =  (N-1)/N \times (N-2)/N \cdots (N-t)/N \times (N-t-1)/N \times 1/N$.
Generalizing this procedure for an arbitrary period $p$, one obtains the values of $S_{0,rm}^{(N)}(t,p)$ displayed in the Table~\ref{table:distr_conj_RM}.
This $N$-order matrix is symmetric and all elements below secundary diagonal are null.

\begin{table}[htb]
\begin{center}
{\tiny
\begin{tabular}{c|c|c|c}
\hline 
$p$ & $t=0$ & $t=1$ & $t=2$ \\
\hline
&&&\\
1 & $\frac{1}{N}$
  & $\frac{N-1}{N} \frac{1}{N}$
  & $\frac{N-1}{N} \frac{N-2}{N} \frac{1}{N}$ \\
&&&\\
\hline
&&&\\
2 & $\frac{N-1}{N} \frac{1}{N}$
  & $\frac{N-1}{N} \frac{N-2}{N} \frac{1}{N}$
  & $\frac{N-1}{N} \frac{N-2}{N} \frac{N-3}{N} \frac{1}{N}$ \\ 
&&&\\
\hline
&&&\\
3 & $\frac{N-1}{N} \frac{N-2}{N} \frac{1}{N}$
  & $\frac{N-1}{N} \frac{N-2}{N} \frac{N-3}{N} \frac{1}{N}$
  & $\frac{N-1}{N} \frac{N-2}{N} \frac{N-3}{N} \frac{N-4}{N} \frac{1}{N}$ \\
&&&\\
\hline
\end{tabular}
}
\end{center}
\caption{Joint distribution for transient time and cycle periods. 
The table is symmetric and the relevant quantity is the number of explored sites $n_e=t+p$.}
\label{table:distr_conj_RM}
\end{table}

With a simple inspection of the values in Table~\ref{table:distr_conj_RM} one concludes  that the transient time and attractor period joint distribution is: 
\begin{eqnarray}
\nonumber
S_{0,rm}^{(N)}(t,p) & = & \frac{1}{N} \prod_{j=1}^{t+p-1} \frac{N-j}{N} \\ 
                    & = & \frac{\Gamma(N)}{\Gamma(N-t-p+1)N^{t+p}} \; ,
\label{eq:distr_conj_RM}
\end{eqnarray}
where $t = 0, 1, 2, \ldots, N-p$. 
Using this expression, one can obtain the marginal probability for the attractor period $p$.
Thus:
$S_{0,rm}^{(N)}(p)  =  \sum_{t=0}^{N-p} S_{0,rm}^{(N)}(t,p) = \sum_{j=p}^N \Gamma(N)/\Gamma(N-j+1)N^j$.
This result agrees the ones obtained in Refs.~\cite{lima_phd,derrida:2:1997}, with $N \gg 1$: $S_{0,rm}^{(N)}(p)  =  1/\sqrt{N} \int_{p/\sqrt{N}}^{\infty} \mbox{d}y \; e^{-y^2/2}  =  \sqrt{\pi/(2N)} \mbox{erfc} (p/\sqrt{2N})$ where $\overline{p} = \sqrt{\pi N/8}$ and  $\sigma_{p}^2 = (2/3- \pi/8) N$.

Notice that Table~\ref{table:distr_conj_RM} symmetry implies to  $ S_{0,rm}^{(N)}(p)  =  S_{0,rm}^{(N)}(t-1)$, i.e., the marginal distributions for the period $p$ and the transient $t$ are identical, when time is retarded by one unity.
Also, observe the strong influence of $N$ in the form of the distribution, which diverges in the thermodynamics limit.

\begin{figure}[htb]
\begin{center}
\includegraphics[angle=-90,width = \columnwidth]{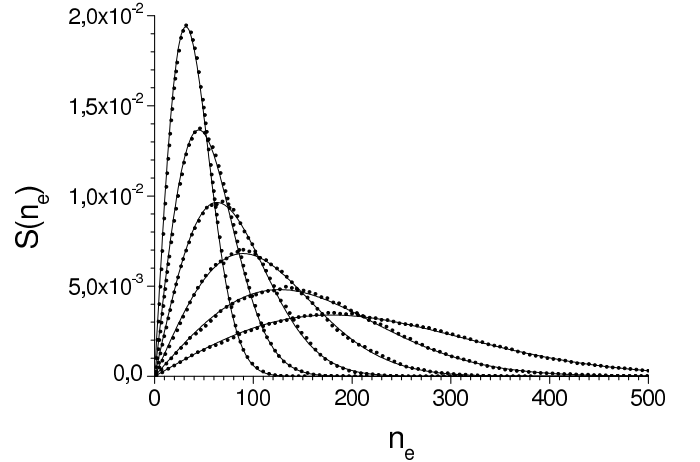}
\caption{Distribution of the number of explored sites given by Eq.~\ref{eq:distr_RM_sit_explor} and numerical simulation. From the top to the bottom, the curves refers respectively to $N=1000$, $2000$, $4000$, $8000$, $16000$, $32000$, $64000$ and $128000$ points by map. 
The greater $N$ is, the wider the distribution becomes.}
\label{fig:RM_Exploracao}
\end{center}
\end{figure}

In terms of the number of explored sites $n_e = t+p = 1,2,\ldots,N$, Eq.~\ref{eq:distr_conj_RM} becomes:
\begin{eqnarray}
\label{eq:distr_RM_sit_explor}
S_{0,rm}^{(N)}(n_e) & = & \frac{n_e\Gamma(N)}{\Gamma(N-n_e+1)N^{n_e}} \; .
\end{eqnarray}
Fig.~\ref{fig:RM_Exploracao} shows the validation of Eq.~\ref{eq:distr_RM_sit_explor} through numerical simulation.

The analogy we have estabilished to the geometric distribution may also be applied to the RMM.
Adopt as a failure the exploration of a new site and as success the revisit to a given  site.
Consider that the walker leaves from site $s_1$.
For the walker to explore the map  in the first time step (visit a new site), site $s_1$ must be connected to some of the $N-1$ other sites, but not to itself. 
The exploration probability in the first step is $q_{0,rm}^{(N)}(1)  = (N-1)/N$.
The walker then goes to the site $s_2$.
To explore a new site in the second step, site $s_2$ must be connected to some of the $N-2$ reminder sites, but neither to itself nor to $s_1$.
Therefore $q_{0,rm}^{(N)}(2) = (N-2)/N$. 
This reasoning may be generalized for an arbitrary time step $j \le N$: 
\begin{equation}
q_{0,rm}^{(N)}(j)  =  1 - \frac{j}{N} \; .
\end{equation}
Thus, the Eq.~\ref{eq:distr_RM_sit_explor} is rewritten in the form: 
\begin{equation}
S_{0,rm}^{(N)}(n_e) = \left[1 - q_{0,rm}^{(N)}(n_e)\right] \prod_{j=1}^{n_e-1} q_{0,rm}^{(N)}(j) \; .
\label{eq:distr_RM_sit_explor_geom}
\end{equation}
The equivalence between the Eqs.~\ref{eq:distr_RM_sit_explor} and~\ref{eq:distr_RM_sit_explor_geom} is immediatly verified.
Notice that the exploration probability decreases in arithmetic progression each step of the trajectory, contrasting to the preceding models, where the subsistence probability decreases in harmonic progression.
Another notable difference is that, in the thermodynamic limit, the exploration probability is unitary.
In this way, the trajectory may have an infinite transient time, what characterize the chaos.

The cumulative distribution for the number of explored site $n_e$ may be obtained (according to the Appendix~\ref{apendice2}):
$ F_{0,rm}^{(N)}(n_e)  = 1 - \Gamma(N)/\Gamma(N-n_e)N^{n_e}$. 

\section{Conclusion}
\label{sec:concl}

We have obtained the joint distribution for the $\mu = 1$ tourist walk for an arbitrary dimension in an euclidean space and also in two mean field models.
These distributions are parameterized by $I_d$ and the number of explored sites $n_e = t + p$.
The former has been first introduced by Cox in the context of spatial statistics.
Except for the RLM, we have not succeeded in obtaining a closed analytical form in the finite size regime for the joint distributions in an arbitrary dimension.
Except for the RMM, these distributions show a fast (factorial) convergence to the attractors. 
It is an open question whether this behavior remains or not valid when the tourist has short  range memory.

\acknowledgements

The authors thank O. Kinouchi and R.N. Onody for interesting discussions on this subject. 
A.S.M. would like to acknowledge the support from the Brazilian agency CNPq (305527/2004-5).

\appendix

\section{Some Special Functions}
\label{apendice1}

Some special functions which are extensively used along the text are reviewed here.   
Initially, consider the gamma function~\cite{stegun_chap} $\Gamma(z)  =  \int_0^{\infty} \mbox{d}t \; t^{z-1} e^{-t}$, which main property $\Gamma(z)  =  (z-1)\Gamma(z-1)$ enables it to be a generalization of the factorial
$\Gamma(z)  =  (z-1)!$.
For $|z| \gg 1$ e $|\mbox{arg} \; z|<\pi$, Stirling's approximation~\cite{stegun_chap} is used $\Gamma(z)  \approx  \sqrt{2\pi/z} (z/e)^z 
[1 + 1/(12z) + 1/(288z^2) + \cdots ]$.  
The non-normalized incomplete gamma function~\cite{stegun_chap} is defined as:
\begin{eqnarray}
\label{eq:gama_inc_n_norm}
\gamma(a,b) & = & \int_0^b \mbox{d}t \; t^{a-1} e^{-t}
\end{eqnarray}
and presents the following property~\cite{stegun_chap} $\gamma(1/2,x)  =  2 \int_0^{\sqrt{x}} \mbox{d}t \; e^{-t^2} = \sqrt{\pi} \; \mbox{erf}(\sqrt{x})$,  where the error function~\cite{stegun_chap} is defined as:
$\mbox{erf}(z)  =  2/\sqrt{\pi} \int_0^z  \mbox{d}t \; e^{-t^2}
= 2/\sqrt{\pi} \sum_{k=0}^\infty (-1)^k z^{2k+1}/[k!(2k+1)]$, which is monotoly increasing from $\mbox{erf}(0)=0$ to $\mbox{erf}(\infty)=1$.

The normalized incomplete beta function~\cite{stegun_chap} is defined as: 
\begin{eqnarray}
\label{eq:beta_incompl}
I_z(a,b) & = & \frac{1}{B(a,b)} \int_0^z \mbox{d}t \; t^{a-1}(1-t)^{b-1}
\end{eqnarray}
with $\mbox{Re}(a)>0$ e $\mbox{Re}(b)>0$ and the beta function~\cite{stegun_chap} is defined as being the normalization factor of $I_z(a,b)$: $B(a,b) = B(b,a)  =  \int_0^1 \mbox{d}t \; t^{a-1}(1-t)^{b-1} = \Gamma(a)\Gamma(b)/\Gamma(a+b)$. 
One can conclude that the beta function is an extension of the inverse of the combination of the Newton's binomial. 

It is also convenient to define the complementary functions: 
The complementar non-normalized gamma function is written as:
$\Gamma(a,b)  =  \int_b^\infty \mbox{d}t \; t^{a-1} e^{-t} = \Gamma(a) - \gamma(a,b)$ 
and the complementar error function~\cite{stegun_chap} is defined by: $\mbox{erfc}(z)  =  2 \int_z^\infty  \mbox{d}t \; e^{-t^2} /\sqrt{\pi}= 1-\mbox{erf}(z)$. 
For $|z| \gg 1$, the complementar error function has the following assymptotic form: 
$\mbox{erfc}(z) \approx  e^{-z^2}/(z \sqrt{\pi})$.

Consider the behavior of Eq.~\ref{eq:beta_incompl} when $b \gg a \sim 1$.
In this case, Eq.~\ref{eq:beta_incompl} is written as:
$ I_z(a,b)  \approx b^a\int_0^z \mbox{d}t \; t^{a-1}(1-t)^b / \Gamma(a)$. 
If $t \ll 1$, then the factor $(1-t)^b = e^{b \ln(1-t)} \approx e^{-bt}$, 
so that: $I_z(a,b)  \approx  \gamma(a,b z)/\Gamma(a)$, where $\gamma(a,b)$ is given by Eq.~\ref{eq:gama_inc_n_norm}. 
When $a = 1/2$ and $b \gg 1$ one has: $I_z(a,b)  \approx  \gamma(1/2,b z)/\Gamma(1/2) \approx \mbox{erf} (\sqrt{b z}) = 1- \mbox{erfc} (\sqrt{b z})$ and 
\begin{equation}
I_z(a,b)  \approx  1 - \frac{1}{\sqrt{\pi}} \; \frac{e^{-b z}}{\sqrt{b z}} \; .
\label{eq:gama_aprox}
\end{equation}

\section{Cumulative distribution in subsistence probability terms}
\label{apendice2}

Here is convinient to deal with the cumulative of the generalized geometric distribution:
\begin{equation}
S(t) = [1-q(t+1)] \prod_{j=1}^t q(j) \; .
\nonumber
\end{equation}
It can be obtained as follow:
\begin{eqnarray*}
S(0) & = &            1 - q(1) \\
S(1) & = &         q(1) - q(1)q(2) \\
S(2) & = &     q(1)q(2) - q(1)q(2)q(3) \\
\vdots &   & \vdots \\
S(t) & = & q(1)q(2) \cdots q(t) - q(1)q(2) \cdots q(t+1) \; .
\end{eqnarray*}
Summing both members, yields:
\begin{equation}
F(t) = \sum_{k=0}^t S(k) = 1 - \prod_{j=1}^{t+1} q(j) \; .
\end{equation}

\bibliographystyle{apsrev}

\end{document}